\documentstyle[12pt]{article}
\newcommand{\be}{\begin{equation}}
\newcommand{\ee}{\end{equation}}
\newcommand{\bea}{\begin{eqnarray}}
\newcommand{\eea}{\end{eqnarray}}
\begin{document}
~\hfill{\footnotesize UICHEP-TH/97-11; IOP-BBSR/97-38}\\
\vskip1.0cm
\begin{center}
{\Large{\bf Exact Solution of a Class of Three-Body  Scattering Problems
in One Dimension}}
\vskip2.5cm
Avinash Khare$^{++}$
\vskip0.5cm
{\it Institute of Physics,\\Sachivalaya Marg,
Bhubaneswar-751 005, India}
\vskip0.5cm
\vskip0.5cm
Uday P. Sukhatme$^{**}$
\vskip0.5cm
{\it Department of Physics (m/c 273), \\ University of Illinois at Chicago,
845 W. Taylor Street, Room 2236 SES, \\ Chicago, Illinois ~ 60607-7059, U.S.A.}
\end{center}
\vskip2.5cm
\begin{abstract}
\end{abstract}
\noindent We present an exact solution of the three-body scattering
problem for a one parameter family of one dimensional potentials
containing the Calogero and Wolfes potentials
as special limiting cases. The result is an interesting nontrivial
relationship between the final momenta $p'_i$ and the
initial momenta $p_i$ of the three particles. We also discuss another one
parameter family of potentials
for all of which $p'_i=-p_i~(i=1,2,3)$.
\vfill
$^{++}$ khare@beta.iopb.stpbh.soft.net\\
$^{**}$ sukhatme@uic.edu
\newpage
\thispagestyle{empty}
More than two decades ago, Marchioro \cite{Mar} gave the complete solution of
both the classical and
quantum mechanical scattering problems of three equal mass
distinguishable particles in one dimension interacting  via the two body
Calogero potential \cite{Cal69}
\be\label{1}
V(x_1, x_2, x_3) = {g\over (x_1-x_2)^2} + c.p.~~~,
\ee
where we have used the notation c.p. to denote the terms obtained
by cyclic permutation of the particle
coordinates $x_1,x_2,x_3$.
It was shown that if $p_i$ and
$p'_i$ (i=1,2,3)
denote the momenta of the incoming and outgoing particles respectively then
there is a surprisingly simple relation between the momenta:
\be\label{2}
p'_1 = p_3 ~~, ~~ p'_2 = p_2 ~~, ~~ p'_3 = p_1~~.
\ee
For the classical case, additional
simple relations between the next-to-leading terms in the
asymptotic expressions for the positions of the particles were also derived.
Four years later, Calogero and
Marchioro \cite{Cam} gave the solution to the classical as well as quantum
mechanical three-body scattering problem when the particles interact
via the three-body Wolfes potential \cite{Wol} given by
\be\label{3}
V(x_1, x_2, x_3) = {3g\over(x_1+x_2-2x_3)^2} + c.p.~~.
\ee
For this interaction, they showed that
\be\label{4}
p'_1 = - p_2 ~~, ~~ p'_2 = - p_1 ~~, ~~ p'_3 = -p_3~~.
\ee

The purpose of this note is to raise and answer the following question: are
there other
potentials for which the three-body scattering problem is exactly solvable,
and if so, what is the relationship between the incoming
and outgoing momenta? We will show that both in the quantum and
classical
cases, if the three equal mass particles are interacting via the three-body
potential
\be\label{5}
V(x_1, x_2, x_3) = ~{g\over [{(x_1-x_2)} \cos \delta
+ {(x_1+x_2-2x_3)\over\sqrt 3} \sin \delta ]^2} + c.p.~~,
\ee
then $p'_i$ and $p_i$ are related by
\be\label{6}
\Bigg ( \matrix{p'_1\cr p'_2\cr p'_3\cr}\Bigg )
= \Bigg (\matrix{0 & -a & b\cr -a & b & 0\cr b & 0 & -a\cr} \Bigg )
\Bigg ( \matrix{p_1\cr p_2\cr p_3\cr}\Bigg )~~~,
\ee
where
\be\label{7}
a = {2\over\sqrt 3} \sin ({2\delta}) ~~, ~~
b = {2\over\sqrt 3} \sin ({\pi\over 3}-{2\delta})~~.
\ee
The coupling constant $g$ is non-zero, and as usual,
we assume $g > -{\hbar^2}/4m$ to prevent collapse.
Here $\delta$ is an arbitrary phase angle such that $0\leq \delta \leq \pi/3$.
The potential (5) is a fairly  complicated, translationally
invariant three-body potential which interpolates smoothly between the
two-body Calogero potential ($\delta = 0$) and the three-body Wolfes
potential ($\delta = \pi/6$). Note that the potential of eq. (5) has a
period $\pi/3$ in the angle $\delta$.
We will take $\delta$ in the range $0 \leq \delta \leq \pi/6$.
For $\pi/6 \leq \delta \leq \pi/3$,
then $p_i$ and $p'_i$ are still related by eqs. (6) and (7) but with $\delta$
replaced by $\delta' = \pi/3 - \delta$.

We also discuss an even more general class of three-body potentials given by
$$V(x_1, x_2, x_3) = {g\over [{(x_1-x_2)} \cos \delta
+ {(x_1+x_2-2x_3)\over\sqrt 3} \sin \delta ]^2 }$$
\be\label{8}
+ {f\over [{(x_1-x_2)} \sin \delta
+ {(x_1+x_2-2x_3)\over\sqrt 3} \cos \delta ]^2} + c.p.~~,
\ee
where the coupling constants $g,f$ are both non-zero and
$0\leq \delta \leq \pi/3$. We show that these potentials
are strictly isospectral, in the sense that, for all
of them, irrespective of the value of $\delta,$ the incoming and outgoing
momenta satisfy
\be\label{8a}
p'_1 = -p_1 ~~, ~~ p'_2 = -p_2 ~~, ~~ p'_3 = - p_3~~.
\ee

At first sight, it is difficult to reconcile the results of eqs. (9)
and (6). One might wonder how the simple result of eq. (9) is
obtained for the potential of eq. (8) for all finite values of the
coupling constant $f$, whereas when $f$ goes from being small to zero,
the result for the outgoing momenta
suddenly and discontinuously
jumps to eq. (6) whereas the potential goes smoothly
to eq. (5). A similar situation also occurs when $g=0$. We would like to
remark that this discontinuous change of results arises and was discussed
in the paper of Calogero and Marchioro [3],
which corresponds to our potentials (5) and (8) for the value $\delta = 0$.
It is also nicely discussed by Marchioro in ref. [1] in the
classical context. This paper contains a figure which shows the
scattering process for two values of the coupling constant. For larger $g$,
the interaction is stronger and the particles have a larger
``distance of closest approach" before backscattering.
For smaller $g$, the interaction is weaker, and the particles come much
closer before backscattering. Indeed, backscattering occurs
no matter how small $g$ is, provided that it is non-zero. [This type of
argument is also true for one-dimensional Rutherford scattering.] However, if
$g=0$, there is no repulsive force, and hence no backscattering, which gives
a heuristic argument for the discontinuous change of results when a coupling
constant is made to vanish.

The proof of all the above results for both the
classical and the quantum cases follows by generalizing the
procedure of refs. \cite{Mar} and \cite{Cam}. Hence, in this paper,
we stress only
those points which are different from the discussion
of refs. \cite{Mar} and \cite{Cam}.

Calogero \cite{Cal69} showed that the scattering problem is best solved in
Jacobi coordinates defined by
\be\label{9}
R = {(x_1+x_2+x_3)\over 3}~,~x = {(x_1-x_2)\over \sqrt 2}~,~
y = {(x_1+x_2-2x_3)\over \sqrt 6}~~.
\ee
Throughout this paper, we work in the center of mass frame and following
\cite{Mar}, for
simplicity, take the origin of the $x$-axis to coincide with the position of
the center of mass of the three-body system so that
\be\label{10}
x_1 + x_2 + x_3 = 0 \, .
\ee
We now define the polar coordinates $(r,\phi)$ by
\be\label{11}
x = r \sin \phi~,~y = r \cos \phi~,~
r^2 = {1\over 3}[(x_1-x_2)^2+(x_2-x_3)^2 + (x_3-x_1)^2]~,
\ee
where the variables $r$ and $\phi$ have the ranges $0 \leq r \leq \infty$
and $0\leq \phi \le 2\pi$. In terms of these coordinates, the relative
Hamiltonian (after the elimination of the center of mass)
corresponding to the potential (\ref{5}) takes the simple form
\be\label{12}
H = -{\hbar^2 \over 2m} ({\partial^2 \over \partial r^2}
+ {1\over r}{\partial \over \partial r}) +{M\over r^2}~~,
\ee
where
\be\label{13}
M = -{\hbar^2 \over 2m}{\partial^2 \over \partial \phi^2}
+ {9g\over 2 \sin^2 (3\phi+3\delta)}~~,
\ee
where $0 \leq \delta \leq \pi/6$. Note that $M$ is unchanged
under the transformation $\delta \longrightarrow \delta +{q\pi}/3$ where
$q = 0,1,...,5$, which just reflects the
periodicity of $V(x_1,x_2,x_3)$.

>From eqs. (\ref{12}) and (\ref{13}) we  find that the problem is effectively
the Calogero problem with $\phi$ replaced by $\phi+\delta$. As in that case,
the singular nature of the interaction disconnects the wave functions (apart
from a symmetry requirement in the case of identical particles) in different
sectors of the configuration space corresponding to different intervals of
values of the angular variable $\phi$. For simplicity, as in [1] and [3] we
shall assume that the particles are distinguishable. Hence we consider the
wave functions that differ from zero only in the sector
\be\label{13a}
-\delta \leq \phi (t) \leq \pi/3 - \delta \, .
\ee
Of course, this may be replaced by any one of the other five sectors by
permuting the particles. Now, in the center of mass frame, using
eq. (\ref{10}) it is easily seen that in terms of polar coordinates,
the positions $x_i$ are given by
\be\label{13b}
x_i = - \sqrt{2\over 3} r \cos [\phi + i({2\pi \over 3})] \, , \ i = 1,2,3 \, .
\ee
It is then not difficult to see that the sector defined by eq. (\ref{13a}) is
characterized by the property
\be\label{13c}
x_1 > x_3 < 0 , x_2 > x_3 , \mid x_1 - x_2 \mid < x_1 - x_3 ,
\mid x_1 - x_2 \mid < x_2 - x_3 \, .
\ee
The scattering problem is now easily discussed by following the steps given in
[3] and replacing $\phi$ by $\phi + \delta$ at appropriate places. In
particular, the eigenfunctions of the angular problem are given by
eq. (2.17c) of [3] with $\phi$ replaced everywhere by $\phi + \delta$.

One major difference from [3] is that one now introduces a symmetry
operation $T$ such that
\be\label{14}
Tr = r, \ T\phi = {\pi\over 3} - \phi - 2\delta~~.
\ee
One can easily show that this $T$ operator when applied to the angular
eigenfunction $\Phi_l$
of the problem (as given by Eq. (2.17c) of [3] but with $\phi$
replaced by $\phi+\delta$) yields
\be\label{15}
T \Phi_l = (-1)^l \Phi_l~~.
\ee
Now following the steps of ref. \cite{Cam}, the relations
(\ref{6}) and (\ref{7})
follow immediately, and furthermore the phase shift is as given
by eq. (2.37c) of
\cite{Cam} which is independent of $\delta$. It is worth noting here that
while deriving these relations in the manner of ref. [3], one has
to place additional
restrictions on the initial momenta $p_i$. In particular, when the ordering
of the particles is as given by eq. (\ref{13c}) then the initial momenta
must satisfy
\be\label{16}
p_1 < p_3 > 0 , p_2 < p_3 , p_3 - p_1 < \mid p_1 - p_2 \mid ,
p_3 - p_2 < \mid p_1 - p_2 \mid \, .
\ee

Let us now discuss the same scattering problem in the classical case. To
exclude collapse, we must now assume that $g \geq 0$. We again work in
the center of mass frame and use the polar coordinates $r$ and $\phi$. In
terms of $r$ and $\phi$, the relative classical Hamiltonian corresponding
to the potential (\ref{5}) is given by
\be\label{17}
H = E = {p_r^2 \over 2m} + {B^2 \over r^2}
\ee
where
\be\label{18}
B^2 = {p_{\phi}^2 \over 2m} + {9g \over {2\sin^2 3(\phi + \delta)}}~~,
\ee
and $E$,$B$ are two constants of motion.
Using the fact that $p_r = m {dr\over dt}, p_{\phi} = mr^2 {d\phi\over dt}$,
 it is easily seen
that the solution to the classical eqs. (\ref{17}) and (\ref{18}) is
\be\label{19}
r(t) = \bigg [ ({2E\over m}) (t-t_0)^2 +{B^2\over E}\bigg ]^{1/2}~~,
\ee
\be\label{20}
\cos 3(\phi (t)+\delta) =
[1-{9g\over 2B^2}]^{1/2} \sin {[ \gamma - 3 \tan^{-1} {(t-t_0)\over \tau}]}~~,
\ee
where $\tau$ = $\sqrt{{m\over 2}} (B/E)$ and
\be\label{21}
\sin \gamma = {{\cos 3 {(\phi(t_0)+\delta)}}\over\sqrt{1-{9g\over
2B^2}}}~~ \, ,~~
\cos \gamma = \pm {{\cos 3 {(\phi(\pm
\infty)+\delta)}}\over\sqrt{1-{9g\over 2B^2}}}~~ \, .
\ee
>From here one immediately obtains
\be\label{22}
\phi (\infty) = {\pi\over 3} -2\delta -\phi (-\infty)~~.
\ee

Let us now discuss the scattering process. Clearly the initial state of the
system is completely specified by the asymptotic expression
\be\label{23}
x_i \stackrel{t\rightarrow -\infty}{\longrightarrow} \ {p_i t \over m}
+a_i, \ i = 1,2,3~~,
\ee
while the final state after the collision is specified by
\be\label{24}
x_i \stackrel{t\rightarrow +\infty}{\longrightarrow} \ {p'_i t \over m}
+a^{'}_i~~.
\ee
On following the steps given in [3] and using eqs. (\ref{19}) to (\ref{21})
it then follows that
$p'_i$ and $p_i$ are related by eqs. (\ref{6}) and (\ref{7}) and further
$a_i^{'}$ and $a_i$ are also related by the same matrix.

Finally let us consider the class of potentials given by eq. (\ref{8}). The
scattering problem for the special case of $\delta = 0$ has already been
discussed \cite{Cam}. In terms
of the polar coordinates $(r,\phi)$ defined by eqs. (\ref{9}) and (\ref{11}),
the Hamiltonian is as given by eq. (\ref{12}) but where
\be\label{25}
M = -{\hbar^2 \over 2m}{\partial^2 \over \partial \phi^2}
+  {9\over 2} \bigg [ {g\over \sin^2(3\phi+3\delta)}
+ {f\over \cos^2(3\phi+3\delta)}\bigg ]~~.
\ee
This $M$ is identical to that of \cite{Cam} except that $\phi$ is replaced
everywhere by $\phi+\delta$ (see their eq. (2.9)).
One can now run through their arguments and show that
irrespective of the value of $\delta$, the outgoing and
incoming momenta
satisfy $p'_i = -p_i$ and further the phase shift in all the cases is as given
by eq. (2.37a) of \cite{Cam}.

Similarly, one can show that the solution to the classical equations is as
given
by eq. (\ref{19}) and
\be\label{26}
\sin 3 (\phi(t)+\delta) = \alpha +
\beta \sin [ \gamma^{(1)}+6 \tan^{-1} {(t-t_0)\over \tau}]~~,
\ee
where $\alpha, \beta, \gamma^{(1)}$ and $\tau$  are given by eqs. (3.9)
to (3.12) of
\cite{Cam}. It immediately follows from here that irrespective of the value of
$\delta, \phi (+\infty) = \phi (-\infty)$, and hence following the steps of
\cite{Cam} one gets
\be\label{27}
p'_i = -p_i~~ , ~~ a'_i = -a_i~~ , ~~ i = 1,2,3~~.
\ee

Thus we have obtained a one continuous parameter family of strictly isospectral
three-body potentials as given by eq. (\ref{8}). Besides, we have also
shown that
the class of
potentials as given by eq. ({\ref{5}) are strictly isospectral to
the potentials
with $\delta$ being replaced by $\delta + q\pi/3$ where $q = 0,1,...,5$.
Besides, they are also strictly isospectral to the potentials with $\delta$
being replaced by $q\pi/3 - \delta$. Thus for any $\delta$ between $0 <
\delta < \pi/6$, there are twelve strictly isospectral potentials while for
$\delta = 0$ and $\delta = \pi/6$ there are six strictly isospectral
potentials.
It is worth noting that if we add the two-body harmonic oscillator
potential $\omega^2 \sum(x_i-x_j)^2$ \cite{Cal69}
to eq. (\ref{8}), these potentials still continue to be strictly isospectral
in the sense that for all of them the spectrum is purely discrete,
independent of $\delta$ and is as given by Wolfes \cite{Wol}.
Further, if we add the two-body harmonic potential
to the class of
potentials given by eq. (\ref{5}), then they are also strictly isospectral
having a purely discrete spectrum which is independent of $\delta$. Note that
this class includes both the Calogero two-body and the Wolfes three-body
potentials.

Finally, it is remarkable that the result for three-body scattering is
rather simple when the interaction is the two-body Calogero
potential as given by eq. (\ref{1}), or the three-body Wolfes potential
as given by eq. (\ref{3}),
or even when the interaction is the sum of the Calogero two-body and
the Wolfes three-body potentials! In this context it is worth recalling that
a simple relationship also holds for a system of $N$-particles
interacting via the two-body Calogero
potential. It is then worth enquiring whether a similar
simple relationship holds good for $N$
particle scattering when they are interacting via the three-body Wolfes
potential or the sum of Wolfes and Calogero potentials. If so, these
will give us additional integrable $N$-body systems in one dimension.

\end{document}